\documentclass[a4paper,12pt]{article}
\usepackage[dvips]{graphicx}

\usepackage{amsmath}

\begin{document}


\title{
Condensation Transitions 
in Nonequilibrium  systems
}
\author{ M.\ R.\ Evans \\[2ex]
        School  of Physics,\\
        The University of Edinburgh,\\
        Mayfield Road,\\
        Edinburgh EH9 3JZ,\\
        U.K.}
\maketitle
\begin{abstract}
Systems driven out of equilibrium can often exhibit behaviour not seen
in systems in thermal equilibrium- for example  phase transitions in
one-dimensional systems. In this talk I will review several
`condensation' transitions that occur when a conserved quantity is
driven through the system. Although the condensation is spatial,
i.e. a finite fraction of the conserved quantity condenses into a
small spatial region, useful comparison can be made with usual
Bose-Einstein condensation. Amongst some one-dimensional examples I
will discuss the `Bus Route Model' where the condensation corresponds
to the clustering together of buses moving along a bus-route. 
\end{abstract}


\section{Introduction}

Broadly speaking, one can consider two types of nonequilibrium
systems: those relaxing towards thermal equilibrium and those held far
from thermal equilibrium {\it e.g.} by the system being driven by some
external field. In the latter case the steady state of the system will
not be described by usual Gibbs-Boltzmann statistical weights rather it
will be  a nonequilibrium steady state. 
A natural way to construct a nonequilibrium steady state is to drive
the system by forcing a current of some conserved quantity, for
example energy or mass, through the system.  Such systems are known as
driven diffusive systems (DDS) \cite{SZ}.

In recent years the
possibility
of  phase transitions and phase separation in
{\em one-dimensional} nonequilibrium systems 
has been explored and some examples are by now well
studied.  To appreciate the significance one should recall the general
dictum that in one-dimensional equilibrium systems phase ordering and
phase transitions do not occur (except in the limit of
zero-temperature, or with long range interaction)
\cite{LL}.

Let us briefly review work on one-dimensional
phase transitions in driven systems.
A very simple one-dimensional  driven diffusive system is the
asymmetric simple exclusion process (ASEP). Here particles hop in a
preferred direction on a one-dimensional lattice with hard-core
exclusion (at most one particle can be at any given site).  Indicating
the presence of a particle by a 1 and an empty site (hole) by 0 the
dynamics comprises the following exchanges at nearest neighbour sites
\begin{eqnarray}
1\ 0 &\to& 0\ 1 \quad\mbox{with rate}\quad 1 \nonumber \\
0\ 1 &\to& 1\ 0 \quad\mbox{with rate}\quad q
\end{eqnarray}

The open system was studied by Krug\cite{Krug91} and boundary
induced phase transitions shown to be possible.  Specifically one
considers a lattice of $N$ sites where at the left boundary site (site
1) a particle is introduced with rate $\alpha$ if that site is
empty, and at the right boundary site (site $N$)
any particle present is removed with rate $\beta$. Thus the dynamical
processes at the boundaries are
\begin{eqnarray}
\mbox{ at site $1$} \quad  0 &\to& 1 \quad\mbox{with rate}\quad \alpha \nonumber \\
\mbox{ at site $N$} \quad  1 &\to& 0 \quad\mbox{with rate}\quad \beta\;.
\end{eqnarray}
These boundary conditions force a steady state current of particles
$J$ through the system. Phase transitions occur when
$\lim_{N{\to}\infty} J$ exhibits non-analyticities.  The steady state
of this system was solved exactly for the totally asymmetric case
\cite{DEHP,SD} and more recently for the general $q$ case
\cite{Sasamoto,BECE}.  When $q<1$ the phase diagram comprises three
phases: a high-density phase where the current is controlled by a low
exit rate $\beta$---one can think of this is queue of cars
at a traffic light that doesn't let many cars through;
a low-density phase where the current controlled by a low injection
rate $\alpha$---think of this as a traffic light that does not let many
cars onto an open road; a
maximal-current phase where both $\alpha,\beta$ are high
($\alpha,\beta > (1-q)/2$) and the
current is $J=(1-q)/4$. Note that since increasing $\alpha$
and $\beta$ doesn't increase the current, the current is saturated.
In the maximal current phase generic
long-range correlations exist, an example being the decay of particle
density from the left boundary to the bulk value $1/2$ which is a
power law $\sim 1/x^{1/2}$ where $x$ is distance from the left
boundary.

On the line $\alpha= \beta < (1-q)/2$ which separates the high and low density
phases one finds coexistence between a region of low density in the
left part of lattice and a region of high density on the right separated by
a `shock' where the density changes sharply over a microscopic distance.

Perodic systems (i.e. a ring of sites)
can also exhibit phase separation when inhomogeneities
or defects are introduced.  A very simple example is to introduce 
into the asymmetric exclusion process a
`slow bond' through which particles hop with a
reduced rate. Then in the steady state one can obtain phase separation
between a region of high density behind the slow bond and a region of
low density in front of the slow bond.  Moving defects ({\it i.e.}
particles with dynamics different from that of the others) have also
been considered and exact solutions obtained \cite{MRE96,KF,Mallick}.
One can think of a slow agricultural vehicle on a country road with a
large queue of cars behind it and open road in front of it.

A further question is whether systems related to the hopping particle
models described so far, but without inhomogeneities, can exhibit
phase ordering. A very simple model was introduced in \cite{EKKM}
comprising three species of conserved particles, amongst which all
possible exchanges are allowed. However a key feature is that the
dynamics has a cyclic symmetry i.e. $A$ particles move preferentially
to the left of $B$ particles which move preferentially to the left of
$C$ particles which in turn move preferentially to the left of $A$
particles.  The model exhibits strong phase separation into pure
domains of $A$ $B$ $C$. Similar strong phase separation occurs
in other related models \cite{LBR}.

A final class of transitions in one-dimensional hopping particle
models is that  involving spatial condensation, whereby a finite
fraction of the particles condenses onto the same site. Examples
include the appearance of a large aggregate
in models of aggregation and fragmentation\cite{MKB} and
the emergence of a single flock in dynamical models
of flocking  \cite{OE}. We will analyse
a simple example of a condensation transition which occurs
in the zero-range process which we now define.

\section{The zero-range process}
\label{Sec:ZRP}
The zero-range process was introduced by Spitzer \cite{Spitzer}
and recent applications and developments have been reviewed in \cite{MRE00}.
We consider a one-dimensional lattice of $M$ sites with sites
labelled $\mu =1 \ldots M$ and periodic boundary conditions
(more generally 
one can consider the zero-range process on a lattice of
arbitrary dimension). Each site
can hold an integer number of indistinguishable particles.  The
configuration of the system is specified by the occupation numbers
$n_{\mu}$ of each site $\mu$.  The total number of
particles is denoted by $L$ and is conserved under the dynamics.
The dynamics of the system is given by
the rates at which a particle leaves a site $\mu$ (one can think of it
as the topmost particle---see Figure 1a) 
and moves to the left nearest neighbour
site $\mu{-}1$. The hopping rates $u(n)$ are a function of $n$ the
number of particles at the site of departure.  Some particular cases
are: if $u(n) =n$ then the dynamics of each particle is independent of
the others; if $u(n) = {\rm const}\quad$ for $n>0$ then the rate at
which a particle leaves a site is unaffected by the number of particles
at the site (as long as it is greater than zero).

The important attribute of the zero-range process is that it has a
steady state described by a product measure. By this it is meant that
the steady state probability $P( \{ n_\mu \})$ of finding the system
in configuration $\{n_1, n_2 \ldots n_M\}$ is given by a product of
factors $f(n_\mu)$ 
\begin{equation}
P( \{ n_\mu \}) = \frac{1}{Z(M,L)} \prod_{\mu=1}^{M} f ( n_{\mu} )\;.
\label{Prob}
\end{equation}
Here the normalisation $Z(M,L)$ is introduced so that the sum of the
probabilities for all configurations,
with the correct number of particles $L$, is one.  

In the basic model described above, $f(n)$ is given by
\begin{eqnarray}
f(n) &=&  \prod_{m=1}^{n} \frac{1}{u(m)}\quad\mbox{for}\quad n\ge 1
\nonumber \\
  &=& 1  \quad\mbox{for}\quad n=0
\label{f1}
\end{eqnarray} 

To prove (\ref{Prob},\ref{f1}) one
simply considers the stationarity condition on the probability of a
configuration (probability current out of the configuration due to
hops is equal to probability current into the configuration
due to hops):
\begin{equation}
\sum_{\mu} 
\theta (n_\mu) u(n_\mu) P(n_1 \ldots n_\mu \ldots n_L) 
=
\sum_{\mu} \theta (n_\mu)
u(n_{\mu{+}1}{+}1) P(n_1 \ldots n_{\mu}{-}1, n_{\mu{+}1}{+}1 \ldots n_L) \;.
\label{station}
\end{equation}
The Heaviside function $\theta(n_\mu)$ highlights that it is the sites
with $n>1$ that allow exit from the configuration (lhs of
(\ref{station})) but also allow entry to the configuration (rhs of
(\ref{station})).  Equating the terms $\mu$ on both sides of
(\ref{station}) and cancelling common factors assuming (\ref{Prob}),
results in
\begin{equation}
u(n_\mu)f(n_{\mu-1}) f(n_{\mu}) = u(n_{\mu+1}+1) f(n_{\mu}-1) f(n_{\mu+1}+1)
\end{equation}
This equality can be recast as
\begin{equation}
u(n_\mu) \frac{f(n_{\mu})}{f(n_{\mu}-1)}
= u(n_{\mu{+}1}+1)\frac{ f(n_{\mu{+}1}+1)}{ f(n_{\mu{+}1})}
=\mbox{constant}
\end{equation}
Setting the  constant equal to unity implies
\begin{equation}
f(n_\mu) =  \frac{f(n_{\mu}-1)}{ u(n_\mu)}
\label{recurr}
\end{equation}
and iterating (\ref{recurr}) leads to (\ref{f1}).
\begin{figure}[htb]
\begin{center}
\includegraphics[scale=0.60]{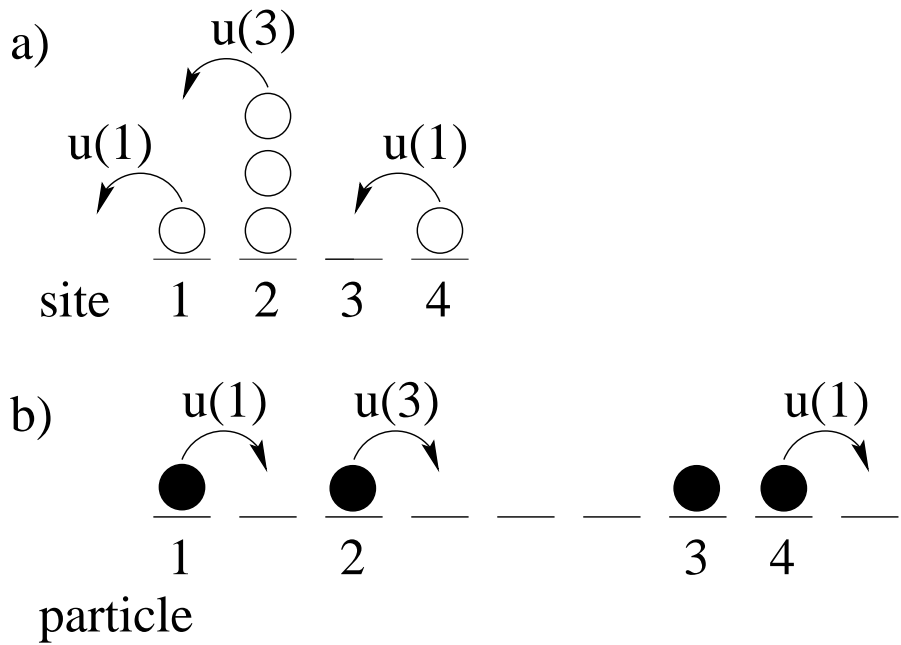}
\end{center}
\caption{\label{fig:ZRP} Equivalence of
zero range process and asymmetric exclusion process.}
\end{figure}

We can easily generalise to  consider an inhomogeneous
system by which we mean the  hopping rates  are site dependent:
the hopping rate out of site $\mu$ when it contains
$n_\mu$ particles is $u_\mu(n_\mu)$.
It is easy to check that the steady state is simply
modified to
\begin{equation}
P( \{ n_\mu \}) = \frac{1}{Z(M,L)} \prod_{\mu=1}^{L} f_{\mu} ( n_{\mu} )
\label{Prob2}
\end{equation}
where $f_\mu$ are given by
\begin{eqnarray}
f_\mu(n) &=& \prod_{m=1}^{n} \frac{1}{u_\mu(m)}
\quad\mbox{for}\quad n\ge 1
\nonumber \\
  &=& 1 \quad\quad \quad\mbox{for}\quad n=0
\label{f2}
\end{eqnarray}
The proof is identical to that  for the homogeneous
case, with the  replacement of $u(n_\mu)$ by $u_\mu(n_\mu)$

There exists an exact mapping from a zero-range process to an
asymmetric exclusion process. This is illustrated in Figure~1.  The
idea is to consider the particles of the zero-range process as the
holes (empty sites) of the exclusion process.  Then the sites of the
zero-range process become the moving particles of the exclusion
process.  Note  that in the exclusion
process we have $M$ particles hopping on a 
lattice of $M+L$ sites.
A hopping
rate in the zero range process
$u(m)$  which is dependent on $m$ corresponds to a hopping rate in
the exclusion process which depends on the gap to the particle in
front. So  the particles can feel each other's presence
and one can have  a long-range interaction.

\section{Condensation Transitions}
\label{Sec:cond}
We now proceed to analyse the steady states of form (\ref{Prob2}) and
the condensation transition that may occur. The important quantity to
consider is the normalisation $Z(M,L)$ as it plays the role of the
partition sum.  The normalisation is defined through the condition
\begin{equation}
Z(M,L) = \sum_{n_1,n_2 \ldots n_M}
\delta(\sum_{\mu} n_\mu{-}L)
 \prod_{\mu=1}^{M} f_\mu( n_\mu )
\label{norm}
\end{equation}
where the $\delta$ function enforces the constraint of $L$ particles.  The
normalisation may be considered as the analogue of a 
canonical partition
function of a thermodynamic system.

We define the `speed' $v$ as the average
hopping rate out of a site
\begin{eqnarray}
v&=& \frac{1}{Z(M,L)}
 \sum_{n_1,n_2 \ldots n_M}
\delta(  \sum_{\mu} n_\mu{-}L)
u(n_1)
 \prod_{\mu=1}^{M} f_\mu( n_\mu )
\nonumber \\
&=&\frac{Z(M,L{-}1)}{Z(M,L)}
\label{speed}
\end{eqnarray}
where we have used (\ref{Prob2},\ref{f2}).  Note that (\ref{speed})
tells us that the speed is independent of site and thus may be
considered a conserved quantity in the steady state of the system.  In
the totally asymmetric system considered in Section~\ref{Sec:ZRP} the
speed is equal to the current of particles flowing between
neighbouring sites.  The
speed is a ratio of partition functions of different system sizes
(\ref{speed}) and corresponds to a fugacity.

We now use the integral representation of the delta function to write
the partition function as
\begin{equation}
Z(M,L) = \oint \, \frac{dz}{2\pi i} \ z^{-(L+1)}\ 
\prod_{\mu=1}^{M} F_\mu(z)\; ,
\label{Zint}
\end{equation}
where 
\begin{equation}
F_\mu(z) = 
 \sum_{m=0}^{\infty} z^{m} \ f_\mu(m)\; .
\label{Fdef}
\end{equation}
For large $M,L$ (\ref{Zint}) is dominated by the saddle point of the
integral and the value of $z$ at the saddle point is the
fugacity. The equation for the saddle point reduces to
\begin{equation}
\frac{L}{M} = \frac{z}{M} \sum_{\mu=1}^{M} \frac{\partial}{\partial z}
\ln F_\mu(z) \label{sad}
\end{equation}
which, defining $\phi= L/M$, can be written as
\begin{equation}
\phi =  \frac{z}{M} \sum_{\mu=1}^{M}  \frac{F'_\mu(z)}{F_\mu(z)}\;.
\label{sad2}
\end{equation}
In the thermodynamic limit,
\begin{equation}
M \rightarrow \infty \;\;\;\mbox{with}\;\;\; L =\phi M\; ,
\label{thermlim}
\end{equation}
where   the density $\phi$ is held fixed,
the question is whether a valid saddle point value
of z can be found from
(\ref{sad2}). We expect that for low $\phi$  the saddle point
is valid but,
as we shall discuss, there exists a maximum value of $z$ 
and if at this maximum value the rhs of (\ref{sad2}) is finite,
then for large $\phi$ (\ref{sad2}) cannot be satisfied.
We now consider how condensation may occur in
the inhomogeneous
and the homogeneous case.

\subsection{Inhomogeneous case}
\label{Sec:Condinhom}
To give an idea of how a condensation
transition may occur we consider the case
$u_\mu(m) = u_\mu$ for $m>0$ {\it i.e.} the hopping rate does not
depend on the number of particles at a site. 
$f_\mu$ is given
by
\begin{equation}
f_\mu(n) = \left( \frac{1}{u_\mu}\right)^{\! n_\mu}
\end{equation}
and the probability of occupancies $\{n_1,n_2,\ldots,n_M\}$ is
\begin{equation}
P(\{ n_1,n_2,\ldots,n_M\}) = \frac{1}{Z(M,L)}
\prod_{\mu =1}^{M}
\left(\frac{1}{u_\mu}\right)^{\! n_\mu}\;.
\end{equation}
The mapping to an ideal Bose gas is evident: the $L$ particles of the
zero-range process are viewed as Bosons which may reside in $M$ states
with energies $E_{\mu}$ determined by the site hopping rates:
$\exp(-\beta E_{\mu}) = 1/u_{\mu}$.  Thus the ground state corresponds
to the site with the lowest hopping rate.  The normalisation $Z(M,L)$
is equivalent to the canonical partition function of the Bose gas.  We
can sum the geometric series (\ref{Fdef}) to obtain $F_\mu$ and
$F'_\mu$ then taking the large $M$ limit allows the sum over $\mu$ to
be written as an integral
\begin{equation}
\phi = \int_{u_{\rm min}}^{\infty} du  {\cal P}(u) \  \frac{z}{u-z}
\label{gce2}
\end{equation}
where ${\cal P}(u)$ is the probability distribution of site hopping rates
with $u_{\rm min}$ the lowest possible site hopping rate.  Interpreting
${\cal P}(u)$ as a density of states, equation (\ref{gce2})
corresponds to the condition that in the grand canonical ensemble of
an ideal Bose gas the number of Bosons per state is $\phi$.  The theory
of Bose condensation tells us that when certain
conditions on the density of low energy states pertain we can have a
condensation transition.  Then (\ref{sad2}) can no longer be satisfied
and we have a condensation of particles into the ground state, which
is here the site with the slowest hopping rate.  

A very simple example is to have just one `slow site'
i.e.
$u_1=p$ while the other $M-1$ sites have hopping rates 
$u_{\mu}=1$ when $ \mu >1$. 
Using the mapping to an exclusion process,
this corresponds  one slow particle
i.e. agricultural vehicle example described earlier.
One can show \cite{MRE96} that
for a high density of particles in the zero range process
(low density of particles in the corresponding asymmetric exclusion process)
we have a condensate since 
site 1 contains a finite fraction
of the particles.
In the low density phase the particles are evenly spread
between all sites.

\subsection{Homogeneous case}
\label{Sec:condhom}
We now consider the homogeneous zero-range process where
the hopping rates $u(n)$ are site independent.
Then (\ref{Fdef}) is independent of $\mu$ and reads
\begin{equation}
F(z) = \sum_{n=0}^\infty   \prod_{m=1}^n \left[ \frac{z}{u(m)} \right]
\label{Fhom}
\end{equation}
The fugacity $z$ must be chosen so that $F$ converges or else we could
not have performed (\ref{Fdef}).  Therefore $z$ is restricted
to $z \leq \beta$ where we
define $\beta$ to be the radius of convergence of $F(z)$.
From (\ref{Fhom}) we see that
$\beta$ is the limiting value  of the
$u(m)$ {\it i.e.} the limiting value
of the  hopping rate out of a site
for a large number of particles at a site.  We interpret
(\ref{sad2}) as giving a relation between the density of holes (number
of holes per site) and the fugacity $z$. 
The saddle point condition (\ref{sad2}) becomes 
\begin{equation}
\label{sad3}
\phi =   \frac{z F'(z)}{F(z)}
\end{equation}
Given that the rhs of (\ref{sad3}) is a monotonically increasing function
of $z$ we deduce that density of
particle increases with fugacity.  However if at $z=\beta$, the
maximum allowed value of $z$, the rhs of (\ref{sad3}) is still finite
then one can no longer solve for the density and one must have a
condensation transition. Physically, the condensation would correspond
to a spontaneous symmetry breaking where one of the sites is
spontaneously selected to hold a finite fraction of the particles.

Thus, for condensation to occur ({\it i.e.} when $\phi$ is large enough
for (\ref{sad3}) not to have a solution for the allowed values of $z$) we
require
\begin{equation}
\lim_{z\to \beta} \frac{F'(z)}{F(z)} < \infty\; .
\end{equation}
We now assume that $u(n)$ decreases uniformly to $\beta$ in the large $n$
limit as
\begin{equation}
u(n) = \beta( 1 + \zeta(n) )
\end{equation}
where $\zeta(n)$ is a monotonically decreasing function.
Analysis of the series 
\begin{eqnarray}
F(\beta) &=& \sum_{n=0}^{\infty}
 \exp \left\{ - \sum_{m=1}^n \ln\left[1+\zeta(m)\right] \right\}
\nonumber \\
F'(\beta) &=& \sum_{n=0}^{\infty} n
 \exp \left\{ - \sum_{m=1}^n \ln\left[1+\zeta(m)\right] \right\}
\label{series}
\end{eqnarray}
reveals that the condition for condensation is simply that
$F'(\beta)$ is finite and this occurs if $u(n)$ decays to $\beta$ more
slowly than $\beta(1+2/n)$.
  (This is easiest to see by expanding $\ln
\left[1+\zeta\right]$ and approximating the sum over $m$ by an
integral in (\ref{series}).)

It is interesting to translate this result
into the language of the
exclusion process.  In this context we can have condensation if
the hop rate of a particle into a gap of size $n$ decays as
$\beta(1+2/n)$ therefore there is an effective long range interaction.

\subsection{Bus route model}
As an example of this let us consider the
`bus route model' \cite{OEC}. 
The model
is defined on a $1d$ lattice. Each site (bus-stop) is
either empty, contains a bus (a conserved particle) or contains a
passenger (non-conserved quantity). The dynamical processes are that
passengers arrives at an empty site with rate $\lambda$; a bus moves
forward to the next stop with rate 1 if that stop is empty; if the
next stop contains passengers the bus moves forward with rate $\beta$
and removes the passengers.

The model thus defined has not been solved but simulations reveal two
regimes.  At high bus density the gaps between buses are evenly
distributed.  However at low bus density there is a condensed regime
where the lead bus has a large gap to the next bus in front of it with
bus-stops full of passengers in between.  The other buses have small
gaps between them.  Thus the buses form a jam of buses and after a
long delay all arrive at a bus-stop at once.

The bus route model can be related to the zero-range process by a
mean-field approximation in which we integrate out the non-conserved
quantity (passengers). The idea is that a bus-stop, next to
bus 1 say, will last have been visited by a bus (bus 2) a mean time
ago of $n/v$ where $n$ is the distance between bus 2 and bus
1 and $v$ is the steady state speed.  
Therefore the mean-field probability that the site next to bus 1
is not occupied by a passenger is $\exp(-\lambda n/v)$. From this
probability an effective hopping rate for a bus into a gap of size $n$
is obtained by averaging the two possible hop rates $1,\beta$:
\begin{equation}
u(n)=\beta+(1-\beta) \exp(-\lambda n/v)\;.
\end{equation}
We can now see that this mean-field approximation to the
bus-route model is equivalent to a homogeneous zero-range process  discussed
earlier.

Since $u(n)$ decays exponentially
the condition for a strict phase
transition in the thermodynamic limit is not met. 
However on any {\em finite} system
for $\lambda$ sufficiently small,
an apparent condensation will be seen. In the bus route
problem this corresponds to the universally irritating
situation of all the buses on the route arriving at once.

\section{Conclusion}
We have shown how the zero range process exhibits two kinds of condensation transition. One is due to having an inhomogeneous system
i.e. we get condensation of particles onto the site with the slowest
hopping rate. Although the condensation is spatial
the mechanism is equivalent to Bose condensation in an
Ideal Bose Gas. 

The other type of condensation occurs on a homogeneous systems and
involves the spontaneous selection of a site onto which a finite
fraction of the particles condense.  Recently this condensation
mechanism has been used to understand the existence or non-existence
of phase separation in a general class of one dimensional driven
systems \cite{KLMST,KLMT}.




\end{document}